\documentclass[twoside,twocolumn,9pt]{article}
\usepackage{extsizes}
\usepackage[super,sort&compress,comma]{natbib} 
\usepackage[version=3]{mhchem}
\usepackage[left=1.5cm, right=1.5cm, top=1.785cm, bottom=2.0cm]{geometry}
\usepackage{balance}
\usepackage{mathptmx}
\usepackage{sectsty}
\usepackage{graphicx} 
\usepackage{lastpage}
\usepackage[format=plain,justification=justified,singlelinecheck=false,font={stretch=1.125,small,sf},labelfont=bf,labelsep=space]{caption}
\usepackage{float}
\usepackage{fancyhdr}
\usepackage{fnpos}
\usepackage[english]{babel}
\addto{\captionsenglish}{%
  
}
\usepackage{array}
\usepackage{droidsans}
\usepackage{charter}
\usepackage[T1]{fontenc}
\usepackage[usenames,dvipsnames]{xcolor}
\usepackage{setspace}
\usepackage[compact]{titlesec}
\usepackage{hyperref}
\usepackage{epstopdf}
\usepackage{lipsum}  
\usepackage{float}
\usepackage{amssymb}
\usepackage{amsbsy}
\usepackage[latin9]{inputenc}

\definecolor{cream}{RGB}{222,217,201}
\graphicspath{{Figures/}}

\begin{document}

\pagestyle{fancy}
\thispagestyle{plain}
\fancypagestyle{plain}{
\renewcommand{\headrulewidth}{0pt}
}
\makeFNbottom
\makeatletter
\renewcommand\LARGE{\@setfontsize\LARGE{15pt}{17}}
\renewcommand\Large{\@setfontsize\Large{12pt}{14}}
\renewcommand\large{\@setfontsize\large{10pt}{12}}
\renewcommand\footnotesize{\@setfontsize\footnotesize{7pt}{10}}
\makeatother

\renewcommand{\thefootnote}{\fnsymbol{footnote}}
\renewcommand\footnoterule{\vspace*{1pt}%
\color{cream}\hrule width 3.5in height 0.4pt \color{black}\vspace*{5pt}} 
\setcounter{secnumdepth}{5}

\makeatletter 
\renewcommand\@biblabel[1]{#1}            
\renewcommand\@makefntext[1]%
{\noindent\makebox[0pt][r]{\@thefnmark\,}#1}
\makeatother 
\renewcommand{\figurename}{\small{Fig.}~}
\sectionfont{\sffamily\Large}
\subsectionfont{\normalsize}
\subsubsectionfont{\bf}
\setstretch{1.125} 
\setlength{\skip\footins}{0.8cm}
\setlength{\footnotesep}{0.25cm}
\setlength{\jot}{10pt}
\titlespacing*{\section}{0pt}{4pt}{4pt}
\titlespacing*{\subsection}{0pt}{15pt}{1pt}
\fancyfoot{}
\fancyfoot[LO,RE]{\vspace{-7.1pt}\includegraphics[height=9pt]{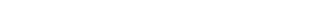}}
\fancyfoot[CO]{\vspace{-7.1pt}\hspace{13.2cm}\includegraphics{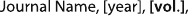}}
\fancyfoot[CE]{\vspace{-7.2pt}\hspace{-14.2cm}\includegraphics{head_foot/RF}}
\fancyfoot[RO]{\footnotesize{\sffamily{1--\pageref{LastPage} ~\textbar  \hspace{2pt}\thepage}}}
\fancyfoot[LE]{\footnotesize{\sffamily{\thepage~\textbar\hspace{3.45cm} 1--\pageref{LastPage}}}}
\fancyhead{}
\renewcommand{\headrulewidth}{0pt} 
\renewcommand{\footrulewidth}{0pt}
\setlength{\arrayrulewidth}{1pt}
\setlength{\columnsep}{6.5mm}
\setlength\bibsep{1pt}
\makeatletter 
\newlength{\figrulesep} 
\setlength{\figrulesep}{0.5\textfloatsep} 
\newcommand{\topfigrule}{\vspace*{-1pt}%
\noindent{\color{cream}\rule[-\figrulesep]{\columnwidth}{1.5pt}} }
\newcommand{\botfigrule}{\vspace*{-2pt}%
\noindent{\color{cream}\rule[\figrulesep]{\columnwidth}{1.5pt}} }
\newcommand{\dblfigrule}{\vspace*{-1pt}%
\noindent{\color{cream}\rule[-\figrulesep]{\textwidth}{1.5pt}} }

\makeatother
\twocolumn[
  \begin{@twocolumnfalse}
{\includegraphics[height=30pt]{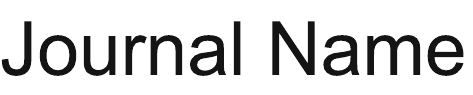}\hfill\raisebox{0pt}[0pt][0pt]{\includegraphics[height=55pt]{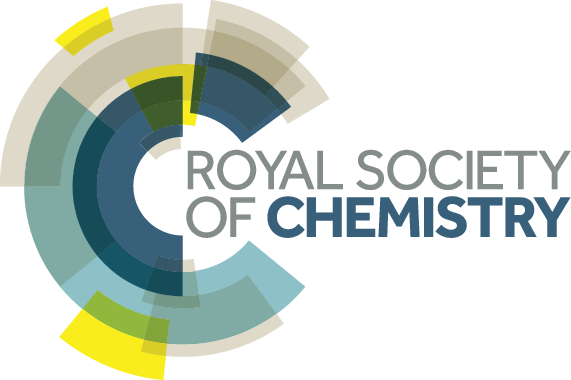}}\\[1ex]
\includegraphics[width=18.5cm]{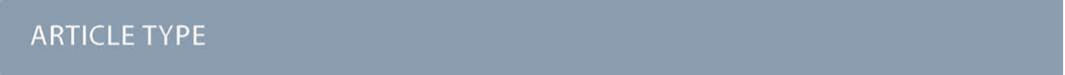}}\par
\vspace{1em}
\sffamily
\begin{tabular}{m{4.5cm} p{13.5cm} }

\includegraphics{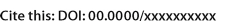} & \noindent\LARGE{\textbf{
Understanding magnetic hyperthermia performance within the ``Brezovich criterion'': beyond the uniaxial anisotropy description
}} \\
\vspace{0.3cm} & \vspace{0.3cm} \\

 & \noindent\large{Daniel Fa\'ilde\textit{$^{a,b}$}, Victor Ocampo-Zalvide\textit{$^{a}$},
 David Serantes$^{\ast}$\textit{$^{a,c}$} and \`Oscar Iglesias$^{\ast}$\textit{$^{d}$}
 } \\

\includegraphics{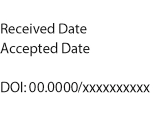} & \noindent\normalsize{
Careful determination of the heating performance of magnetic nanoparticles under AC fields is critical for magnetic hyperthermia applications. However, most interpretations of experimental data are based on the uniaxial anisotropy approximation, which in first instance can be correlated with particle aspect ratio. This is to say, the intrinsic magnetocrystalline anisotropy is discarded, under the assumption that the shape contribution dominates. We show in this work that such premise, generally valid for large field amplitudes, does not hold for describing hyperthermia experiments carried out under small field values. Specifically, given its relevance for \textit{in vivo} applications, we focus our analysis on the so-called "Brezovich criterion", $H\cdot{f}=4.85\times{10^8}A/m\cdot{s}$. By means of a computational model, we show that the intrinsic magnetocrystalline anisotropy plays a critical role in defining the heat output, determining also the role of shape and aspect ratio of the particles on the SLP. Our results indicate that even small deviations from spherical shape have an important impact in optimizing the heating performance. The influence of interparticle interactions on the dissipated heat is also evaluated.
Our results call, therefore, for an improvement in the theoretical models used to interpret magnetic hyperthermia performance.
 } \\

\end{tabular}

 \end{@twocolumnfalse} \vspace{0.6cm}

  ]
\renewcommand*\rmdefault{bch}\normalfont\upshape
\rmfamily
\section*{}
\vspace{-1cm}
\footnotetext{\textit{$^{a}$~Applied Physics Department, Universidade de Santiago de Compostela, 15782 Santiago de Compostela, Spain.
}}
\footnotetext{\textit{$^{b}$~Galicia Supercomputing Center (CESGA), Santiago de Compostela, Spain.
}}
\footnotetext{\textit{$^{c}$~Instituto de Materiais (iMATUS), Universidade de Santiago de Compostela, 15782 Santiago de Compostela, Spain. E-mail: david.serantes@usc.gal
}}
\footnotetext{\textit{$^{d}$~Departament de F\'isica de la Mat\`eria Condensada  and Institut de  Nanoci\`encia i Nanotecnologia Universitat de Barcelona (IN2UB), Mart\'i i Franqu\`es 1, 08028 Barcelona, Spain, Universitat de Barcelona. E-mail: oscariglesias@ub.edu}}




\section{Introduction}\label{intro}
Magnetic hyperthermia cancer treatment with magnetic nanoparticles (NP) received intense research attention in the last decades based on its huge potential for cancer treatment, specially for aggressive tumour types as brain glioblastoma or pancreas, of poor prognosis with usual techniques \cite{tian2021phospholipid}. However, despite the early success in the treatment of patients with recurrent glioblastoma multiforme \cite{maier2011efficacy}, even reaching regulatory approval in the EU \cite{mahmoudi2018magnetic}, the fact is that magnetic hyperthermia did not reach the generated expectations as a promising tool for cancer treatment \cite{dutz2014magnetic,rubia2021whither}. A diversity of reasons lays behind such poor success, ranging from biocompatibility and toxicity concerns of the particles within the biological media \cite{portilla2022different}, to the common problem of nanomedicine approaches to reach significant doses within the desired target \cite{wilhelm2016analysis}. Clearly, several factors need to be improved in order to improve the success of magnetic hyperthermia \cite{beola2019roadmap}. The aim of this work is to study a generally overseen aspect in magnetic hyperthermia studies: the applicability of the usual \textit{effective} uniaxial anisotropy approximation to describe the heating performance of magnetic NP under field conditions suitable for \textit{in vivo} applications. 

The magnetic anisotropy is the key parameter determining the heat released by the NP under the AC field \cite{dennis2013physics}, as it stands for the coupling of the magnetic moment to the lattice. In fact, it defines both the achievable heat, and the performance under a given field amplitude; see details of this double role in Section \ref{model}. Thus, the anisotropy defines the conversion of the absorbed electromagnetic energy -usually described in terms of the \textit{specific loss power}, SLP- into heat \cite{Munoz_PCCP_2017,Martinez-Boubeta_SciRep2013}. For a magnetic field of amplitude $H_{max}$ and frequency $f$, $SLP=HL\cdot{f}$, where $HL$ stands for the hysteresis losses, which can be evaluated as the area of the $M(H)$ loop. Please note that while the related specific absorption rate (SAR) parameter is widely used in materials' science focused hyperthermia studies, SAR has a different meaning in the medical field and thus the SLP parameter should be preferably used \cite{soetaert2017experimental}. 

The problem is that defining the magnetic anisotropy of the NP is not a simple task, as different sources may strongly define the orientation of the particle magnetisation. In addition to the intrinsic (material-defined) magnetocrystalline term, the particle shape \cite{Martinez-Boubeta_SciRep2013,moreno2020role} and surface \cite{garanin2003surface,iglesias2004role} may also strongly influence the behaviour of the particle magnetisation. Furthermore, combination of material properties also leads to very different particle behaviours, for example by creating core/shell morphologies \cite{iglesias2008exchange,khurshid2015anisotropy}, or by fine-tune doping \cite{del2019flower,sathya2016co}. Given the complexity of the problem, and the difficulties in theoretically interpreting data with competing anisotropies, in the literature of magnetic hyperthermia it is often interpreted experimental data in terms of an \textit{effective} uniaxial anisotropy \cite{serantes2014multiplying,conde2015single,simeonidis2016situ}. Such simplification is based on the assumption that for iron oxides (most common system for hyperthermia applications \cite{soetaert2020cancer}) the particle shape is the key contribution to the anisotropy, dominating over the intrinsic magnetocrystalline one \cite{usov2010low,Vallejo-Fernandez_APL_2013}.

Recent theoretical works have suggested, however, that the contribution of this intrinsic magnetocrystalline term is not only non-negligible, but may play a critical role \cite{Usov_Beilstein2019,Gavilan_Nanoscale_2021}. Thus, when considering the intrinsic magnetocrystalline contribution in addition to a shape uniaxial one (i.e. a more realistic approach), it has been shown that in the small-field range the \textit{effective} uniaxial-only approximation deviates significantly from the more realistic case \cite{Gavilan_Nanoscale_2021}. Given the importance of the small fields range for \textit{in vivo} applications, such results suggest that further investigation on the applicability of the \textit{effective} uniaxial approximation is highly necessary. We will show in this work that while the uniaxial-only approximation can reasonably describe large fields and frequencies, it may be completely off for \textit{in vivo} field conditions. Given the complexity of the problem, we have used a computational technique that allowed us to accurately control the different parameters governing the heating performance. 

The work is organised as follows. In Section \ref{model}, the details of the physical model are described, together with the implications regarding heating and the Brezovich criterion. In Section \ref{Sec_Simulation}, the details of the computational model are provided. The results are described in Section \ref{results}, including general field-dependence considerations, specific details regarding SLP under the Brezovich criterion, and the role of interparticle interactions. The conclusions of the work are summarised in Section \ref{conclusions}.

\section{Physical model}\label{model}
The physical model corresponds to the \textit{macrospin} approximation, i.e. the NPs are small enough that their magnetisation is dominated by the exchange energy, leading to coherent rotation of the atomic magnetic moments. Thus, each $i$ particle can be characterised by its magnetic \textit{supermoment} $\vec{\mu}_i$ of magnitude $\mu_i=M_sV$, being $M_S$ the saturation magnetization and $V$ the particle volume. In the current work we will consider magnetite nanoparticles, thus $M_S= 4.8\times 10^5$ A/m.

For ideal non-interacting conditions, the hysteresis behaviour (i.e. heating capability) of each magnetic moment under the AC field is, in first instance, defined by the particle anisotropy energy, as the responsible for creating the local energy barriers that cause irreversible behaviour \cite{dutz2014magnetic}. Furthermore, from the point of view of heating efficiency, it can be said that the anisotropy plays a double role \cite{Munoz_PCCP_2017}: on the one hand, it defines the maximum energy that can be dissipated; on the other hand, it defines the performance under a given field amplitude. Thus, considering magnetic NP with uniaxial anisotropy constant $K$ and volume $V$, the theoretical maximum energy losses $HL_{max}$ would be $8KV$ for easy axes parallel to the field \cite{dennis2013physics}, and $\approx{2KV}$ for a randomly distributed system \cite{munoz2015role}. However, significant heat can only be reached if a a minimum field threshold is overcome; namely, it is required $H_{max}\gtrsim{0.5H_K}$ for significant heating \cite{hergt2006magnetic} (for random easy axes distributions). The $H_K$ value is the so-called anisotropy field, defined as $H_K=2K/{\mu_0M_S}$, where $\mu_0$ is the vacuum permeability . Given its critical role, intense research attention has been devoted to study the role of the anisotropy on heating efficiency \cite{sohn2010optimization,Martinez-Boubeta_SciRep2013,Vallejo-Fernandez2013,nemati2016enhanced,mcghie2017measurement,Mamiya_ACSNano2020,valdes2021role}. 

\begin{figure}[htbp]
   \centering
    \includegraphics[width = \columnwidth]{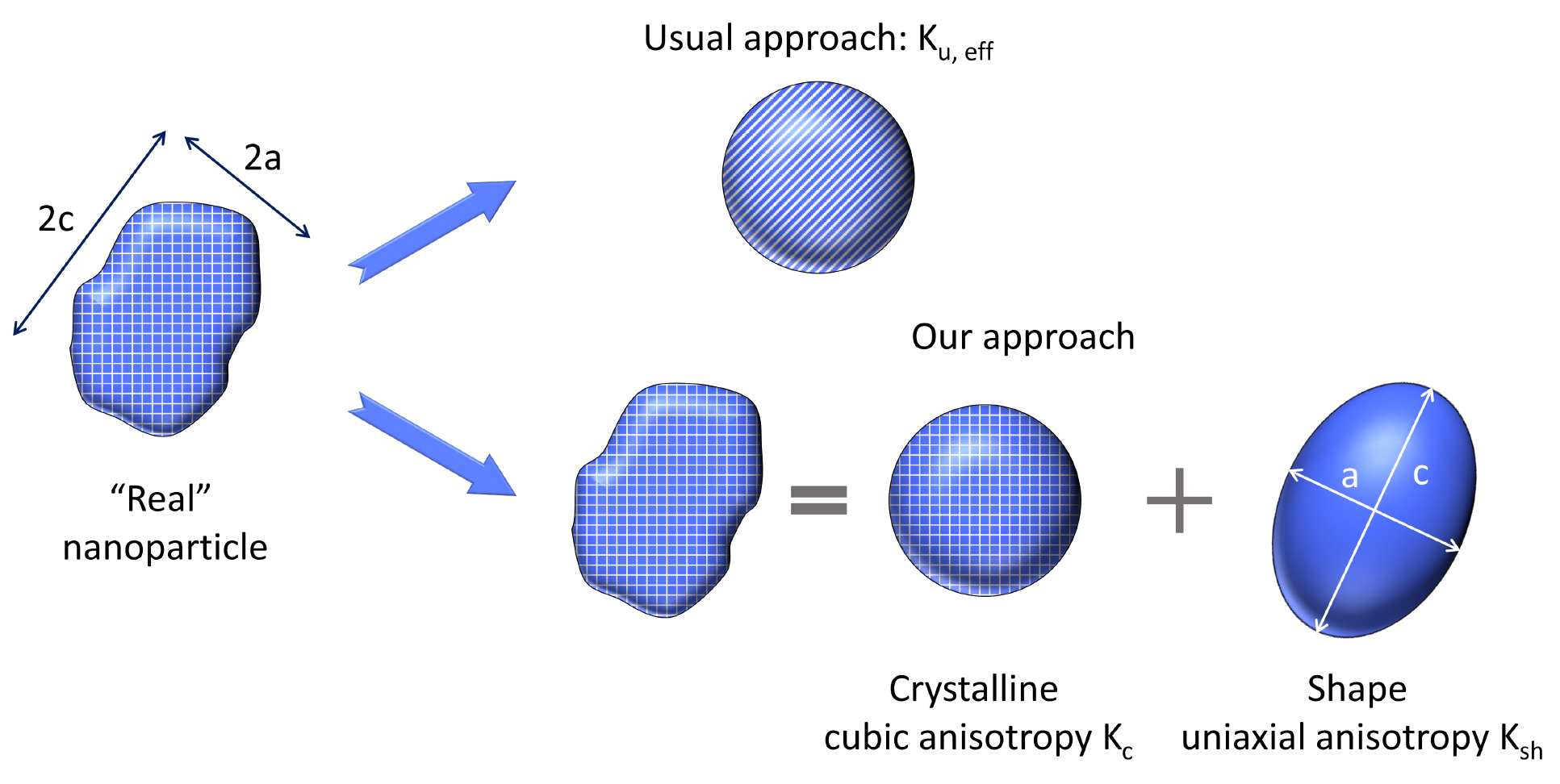}
    \caption{Scheme illustrating the difference between the usual uniaxial-only approximation, and the more realistic approach considered in this work.
		} 
    \label{Fig_Scheme_Shapes}
 \end{figure}

The previous uniaxial-anisotropy description implicitly assumes that the particle has an elongated shape that brings into play a shape anisotropy energy density, $K_{sh}$, that dominates over the intrinsic magnetocrystalline energy density \cite{usov2010low,Vallejo-Fernandez2013}, which for magnetite is cubic and negative, $K_c=-1.1\times 10^4$ Jm$^{-3}$. This is to say, the above description considers only the magnetization and ignores magnetorystalline anisotropy. This is a crucial aspect since, as described above, the anisotropy defines the energy barriers responsible for the irreversible path that results in released energy. For the $K_C<0$ case of the magnetite, the energy barrier $E_B$ is reduced by a factor of $12$ as compared to the case of uniaxial anisotropy \cite{yanes2007effective}, thus allowing much smaller energy to be dissipated, but also lowering the required minimum field amplitude to reach significant heating. The key point of the present work is to study which is the role of this intrinsic magnetocrystalline anisotropy on the behaviour of the heating performance of the particles, under suitable field conditions for \textit{in vivo} applications. A na\"{\i}ve illustration of the difference between the common uniaxial-only model and the "realistic" one used in this work is depicted in Fig. \ref{Fig_Scheme_Shapes}.

In Fig. \ref{Fig_Scheme_Shapes} it is illustrated the difference between the usual approximation and ours: while the usual approach describes a "real" nanoparticle (which is never perfectly spherical) solely in terms of an \textit{effective} uniaxial anisotropy constant, $K_{u,eff}$, our approach maintains the intrinsic magnetocrystalline contribution $K_C$ (which is always present), and considers the sphericity through an additional uniaxial shape anisotropy term, $K_{sh}$, approximated through the aspect ratio of an ellipsoid; see details in Section \ref{Sec_Simulation}.

At this point is very important to highlight that safe \textit{in vivo} applications impose limitations in the field properties. It was first discussed by Brezovich and collaborators \cite{atkinson1984usable,brezovich1988low} that to avoid "discomfort" in patients, the product of field frequency ($f$) and amplitude ($H_{max}$) should 
meet the condition $H_{max}\cdot{f}=4.85\cdot{10^8}$ A/m$\cdot$s (the so-called the "Brezovich criterion"). From a pure materials science perspective, however, such limitation is usually overlooked, being the hyperthermia characterisation often carried out under much higher $H_{max}\cdot{f}$ values. While it has been suggested the possibility of surpassing this limit without causing damage to the patient \cite{herrero2022proposal,kwok2023nonspecific,Rodrigo_IJH2020}, given its historical relevance we have focused our analysis on field conditions matching the Brezovich criterion.

\section{Simulation details}\label{Sec_Simulation}

In order to model the NP ensembles, we consider magnetite  particles with sizes below the single domain limit and, as mentioned in Sec. \ref{model}, represent them by macrospins having a magnetic supermoment $\vec{\mu}_i$. The NPs have randomly distributed easy axes with cubic anisotropy constant $K_c=-1.1\times 10^4$ Jm$^{-3}$; if an additional uniaxial term (see below) is included, the corresponding easy axes are also distributed at random, and uncorrelated from the cubic ones. Typically we simulated a system of $1000$ macrospins, unless otherwise stated.

The simulations were performed as in previous works \cite{Gavilan_Nanoscale_2021,serantes2021nanoparticle} using the OOMMF software \cite{OOMMF} to track the temporal evolution of the system of macrospins under the influence of a sinusoidal alternating field of maximum amplitude $H$ and frequency $f$. The dynamical evolution of each macrospin is described by the Landau-Lifschitz-Gilbert equation, with a random field to account for a finite temperature $T= 300$ K \cite{lemcke2002thetaevolve}. From the simulated hysteresis loops, we obtained the HL value by averaging over different simulation runs and the SLP in W/g as $SLP=HL\cdot{f}/\rho$, where $\rho=5170$ kg/m$^{3}$ is the mass density of magnetite.

\begin{figure}[tbp]
   \centering
    \includegraphics[width = \columnwidth]{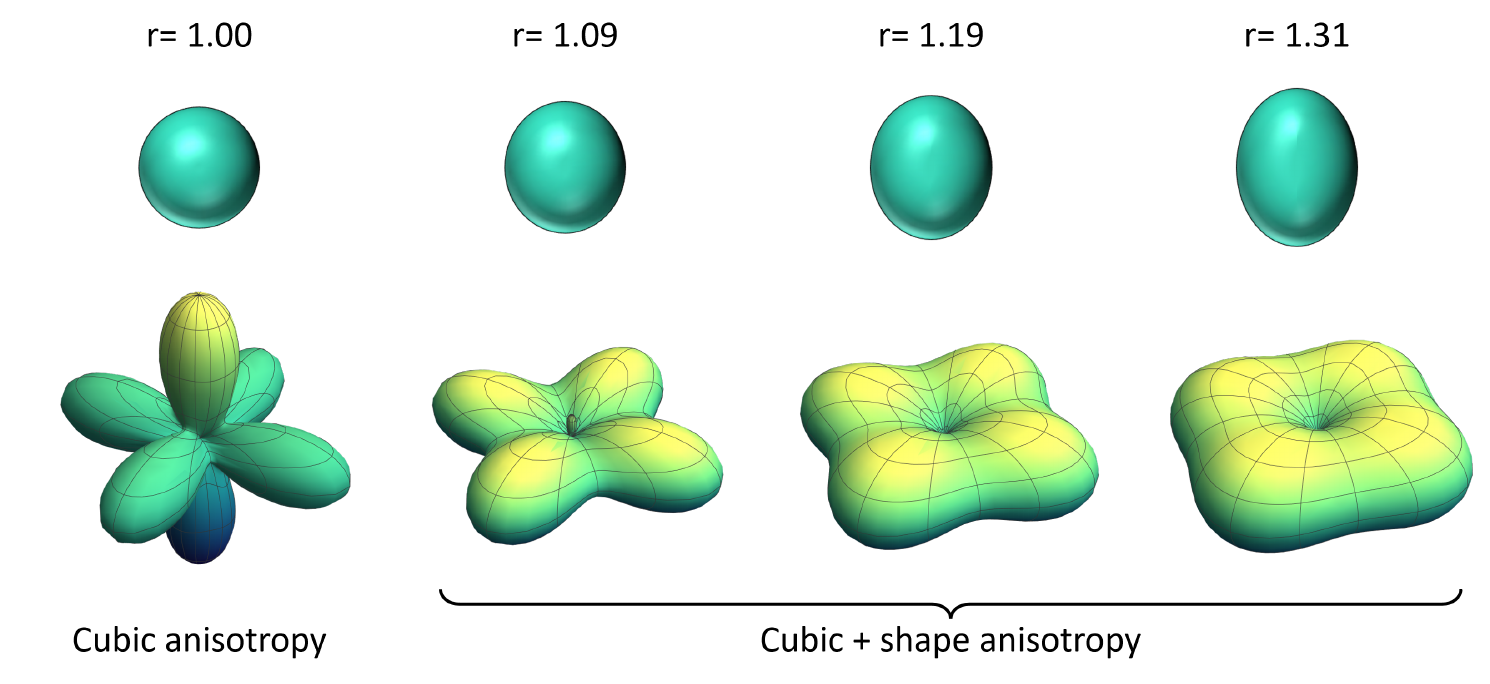}
    \caption{Schematic figure showing our approach to incorporate real particles shape effects. Upper panels show the particle shape with the associated aspect ratio $r$. Lower panels show the anisotropy energy surfaces in spherical coordinates obtained by adding a uniaxial contribution given by Eq. \ref{eq:Ku} to the crystal cubic anisotropy of $K_{C}= -11$ kJ/m$^3$.   
		} 
    \label{Fig_Cubic+Shape}
 \end{figure}

To incorporate the deviation of particle shape from an ideal sphere of diameter $D$ we consider an additional shape anisotropy corresponding to the demagnetizing energy. For a prolate ellipsoid of long axis $c$ and short axes $b=a$, this energy can be written as uniaxial \cite{bertotti1998hysteresis}, with the energy density constant $K_{sh}$ given by
\begin{equation}
K_{sh}=\frac{\mu_0}{2} (N_a-N_c)M^2_s \ ,
\label{eq:Ku}
\end{equation}
where $N_c, N_a$ are the demagnetizing factors along the long and short axes of the ellipsoid. They can be calculated with the expressions \cite{Osborn_PhysRev1945,Stoner_PhilMag1945}
\begin{equation}
N_c=\frac{1}{r^2-1}\left[\frac{r}{\sqrt{r^2-1}} \ln{(r+\sqrt{r^2-1})}-1
\right]   \quad N_a=\frac{1-N_c}{2} \ ,
\label{eq:Nc}
\end{equation} 
where $r=c/a$ is the aspect ratio.
In Fig. \ref{Fig_Cubic+Shape}, we have depicted some representative examples of the anisotropy energy surfaces of the cases we have simulated, together with the shape of particles with different aspect ratios.
The figure illustrates how a spherical NP with only cubic negative anisotropy ($r=1.0$) has easy axes directed along the diagonal directions, but these progressively become redirected towards the $z$ axis as the uniaxial shape anisotropy increases when elongating the NP.
Note that for illustrative purposes we chose this case with the same easy axes directions for cubic and uniaxial anisotropies but, in general, those will be always uncorrelated in the simulations.

In general, we have focused our study on non-interacting particles, the logical first step to understand the complex system under study. In that case, in the simulations we can identify each particle as the basic cubic cell discretization, setting the volume $L^3$ equal to the NP volume $V$. Therefore, given a cell size $L$, the diameter of the sphere having the same volume is $D=1.24 L$. However, given its importance for the application (particles tend to agglomerate when internalized within the cells \cite{Beola_ACSAppl2020}), we have also considered the role of interparticle interactions. In this second case, to vary the sample volume concentration, $c$, we generated assemblies of randomly distributed NPs by adapting the simulation atlas size ($L_{atlas}$) so that  $c=NV/L_{atlas}^3$, being $N$ the number of NPs. In OOMMF this is done by setting the saturation magnetization of the cells not assigned to a particle to $M_s=0$. In all cases, the minimum amount of particles considered in the simulations was $N=1000$. The dipolar interactions between macrospins were taken into account through the demagnetizing energies between simulation cells by switching on the class module \texttt{Ox\_Demag}. In general, the size of the plotted points data for the non-interacting results is bigger than the error bars. However, that is not the case for the interacting conditions, and in those cases the results are averaged over 4 different runs.

\section{Results and discussion}\label{results}

Before analysing in detail the specific aspect of heating performance defined by the different anisotropy contributions, we will first revise the general picture. Thus, in Subsection \ref{subsect-field_dependenceSLP} we analyze the heating performance (in terms of the SLP) as a function of field amplitude, $H_{max}$, for the usual anisotropy-only approach, and the combined cubic plus uniaxial one. Such is a very convenient procedure from the theoretical viewpoint, as the double role of the anisotropy is emphasized through the characteristic sigmoidal shape \cite{iacovita2019hyperthermia,Castellanos-Rubio2021Nov,ovejero2021selective} corresponding to the minor-to-major loop transition \cite{verde2012field}, where the maximum heat at large fields is proportional to the anisotropy \cite{Munoz_PCCP_2017}. Then, in Subsection \ref{subsect SLP for Brezovich conditions} we focus specifically on field/frequency combinations corresponding to the Brezovich criterion. For simplicity, this analysis is at first carried out for non-interacting conditions, so that the role of the different anisotropy contributions is not counteracted by interparticle interactions \cite{conde2015single}. Nevertheless, as previously mentioned, interacting conditions are very relevant for the application. Therefore, in Subsection \ref{subsect interactions} we study the effect of interparticle interactions on SLP.

\begin{figure}[thbp]
   \centering
    \includegraphics[width = \columnwidth]{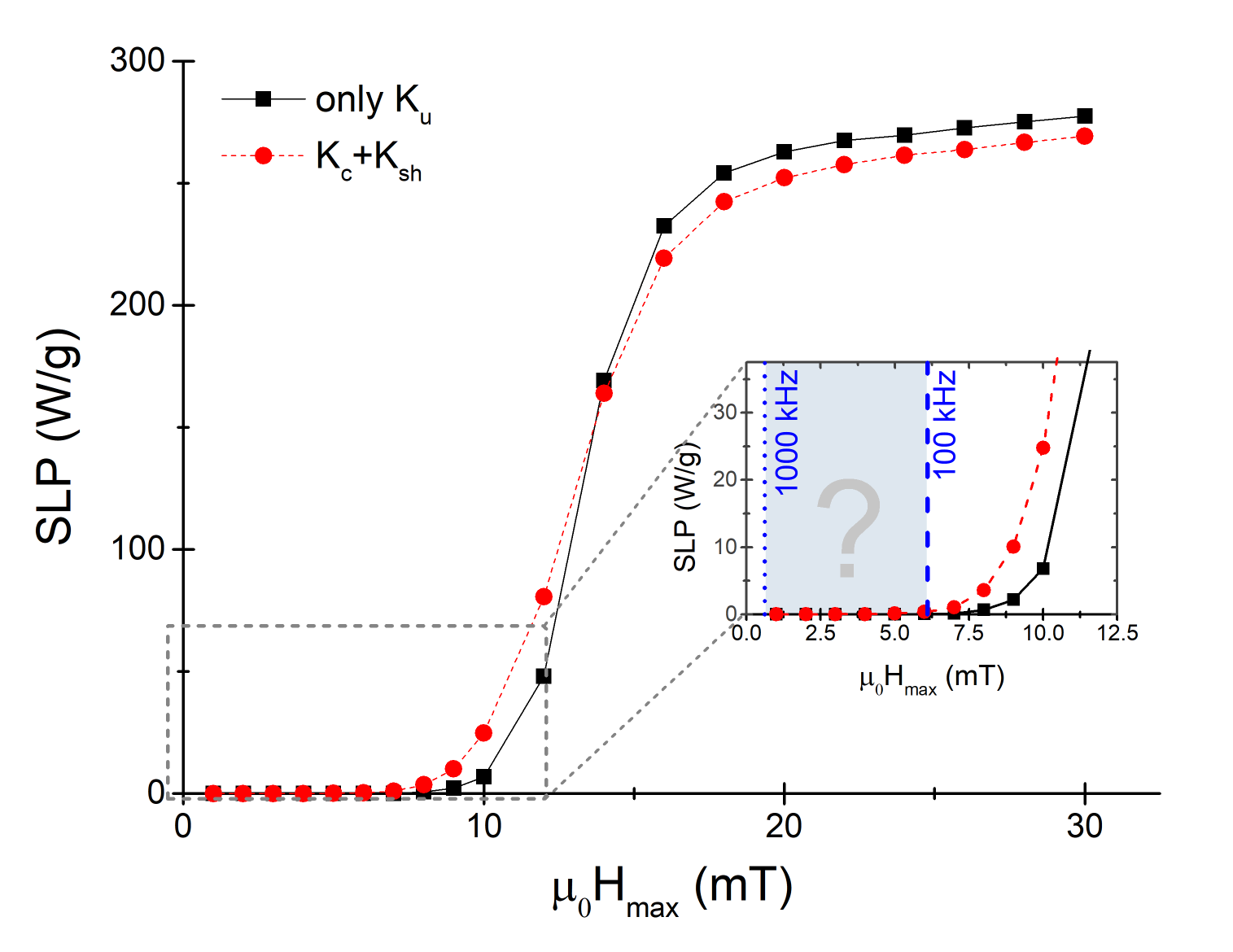}
    \caption{SLP dependence on the maximum AC field $H_{max}$ and frequency $f= 100$ kHz for a spherical particle of $24.8$ nm with only uniaxial anisotropy $K_u=1.1\times 10^4$ J/m$^3$ only, black squares) and for an ellipsoidal particle with cubic anisotropy $K_c= -1.1\times 10^4$ J/m $^3$ and an additional uniaxial contribution $K_{sh}$ of the same value (red circles). Inset: zoom of the low field region where the vertical dashed lines indicate the field values for which the Brezovich criterion is accomplished at the given frequencies.} 
    \label{Fig_SLP_vs_Hmax_v2}
 \end{figure}
\subsection{Field dependence of SLP}\label{subsect-field_dependenceSLP}

Let us first consider a non-interacting assembly of spherical particles and assume they have only uniaxial anisotropy ("only K$_u$" case), as in many models and experimental studies. As a representative size, we started choosing $D\approx{25}$ nm (discretization cell size $L= 20$ nm), as it is interesting for the application point of view and, furthermore, we have previously obtained a rich behaviour if considering also the magnetocrystalline term \cite{Gavilan_Nanoscale_2021}. For the uniaxial anisotropy constant, we chose a usual magnetite-like value, $K_u=1.1\cdot{10^4}$ J/m$^3$. In order to study the range of magnetic fields necessary to induce appreciable heating, we simulated hysteresis loops with increasing values of the maximum ac field $H_{max}$, to obtain a complete SLP \textit{vs.} $H_{max}$ curve. For the sake of generality, we did this systematically for two values of the frequency $f=100$ kHz, and $f=1000$ kHz, so that we covered the usual frequency range in magnetic hyperthermia. The results for the $f=100$ kHz case are shown in Fig. \ref{Fig_SLP_vs_Hmax_v2},

In Fig. \ref{Fig_SLP_vs_Hmax_v2}, the SLP shows a sigmoidal field dependence (black squares), increasing abruptly for fields higher than $10$ mT, and showing a tendency towards saturation for fields higher than the anisotropy field $H_k$. This corresponds to a transition from minor to major loop conditions. However, if the spherical particles are assumed to have only cubic anisotropy with $K_c= -1.1\times 10^4$ J/m$^3$ instead of uniaxial, 
the obtained SLP is negligible for all $H_{max}$ and range of frequencies considered. Therefore, data have not been included in the figure. 

If now a shape anisotropy contribution with the same value as the cubic one ($K_{sh}=1.1\times 10^4$ J/m$^3$) is added to include a change of the shape from spherical to ellipsoidal (i.e. the particles have now combined $K_c+K_{sh}$), the general qualitative of SLP does not change (red circles in Fig. \ref{Fig_SLP_vs_Hmax_v2}). Only a slight increase (decrease) of the SLP with respect to the uniaxial case is obtained at low (high) fields. Note that the increase at low fields can be explained by the appearance of small energy barriers due to the magnetocrystalline term (which can be overcome by smaller fields). Similarly, the decrease at high fields can be interpreted as the result of the appearance of easier reversal paths (smaller energy barriers) again due to the presence of the magnetocrystalline anisotropy.

Since SLP is proportional to $f$, an increase in frequency up to $1000$ kHz just increases the SLP by an order of magnitude with values approaching the SW limit, without changing the qualitative behavior (to make the figure clearer, these results have not been included). This is the usual sigmoidal variation of SLP with $H_{max}$ reported in most models of NP assemblies when studying magnetization dynamics in the macrospin approximation, also with the same effect of such changes with frequency \cite{ovejero2021selective}. 

However, notice that the $H_{max}$ values that give appreciable SLP are beyond the physiological limits imposed by the Brezovich criterion even for the lowest frequency. Considering that in most experimental set-ups $f$ can be varied within the range $100-1000$ kHz, if physiological limits are to be respected, the maximum allowed fields will range from $6.1$ mT to $0.61$ mT as indicated by vertical dashed blue lines in the zoomed in region of Fig. \ref{Fig_SLP_vs_Hmax_v2}. 
These fields are too low to induce any heating, at least for the particle size considered here, that is typical of particles studied experimentally. 
Therefore, an important conclusion of this observation is that the approximation of considering uniaxial anisotropy is invalid to explain why heating occurs under the Brezovich criteria conditions. But then, what is the reason why appreciable heating is observed experimentally?
In what follows, we will try to shed some light on this issue.

\begin{figure}[thpb]
   \centering
    \includegraphics[width = \columnwidth]{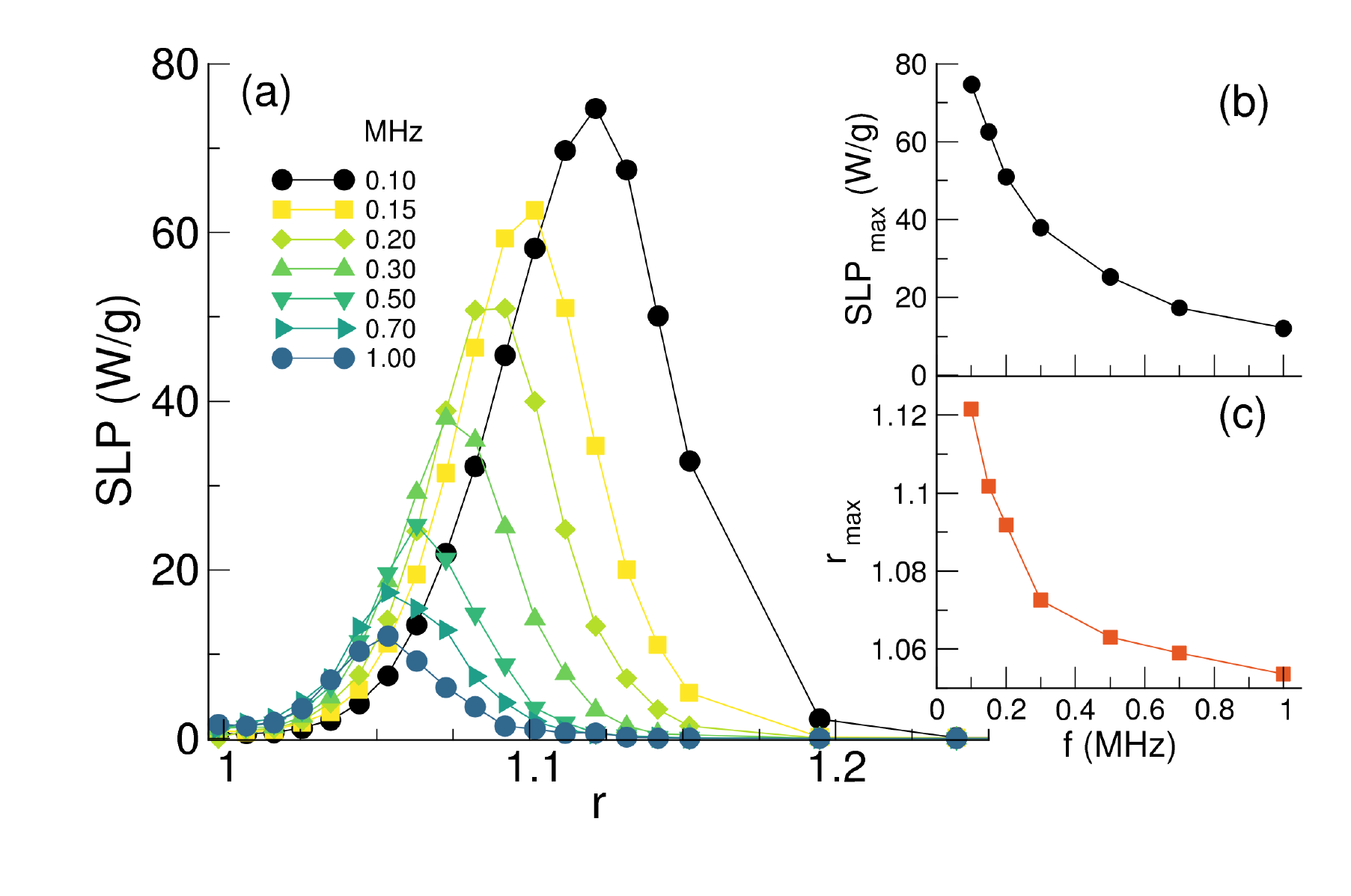}
    \caption{(a) Dependence of SLP on the nanoparticle aspect ratio $r$ for different frequencies for particles of size $D \approx{25}$ nm. (b) Corresponding frequency dependence of the maximum SLP and (c) $r$ at the maximum.}
    \label{Fig_SAR_Bre_r_f's_Victor_Inset}
 \end{figure}

\subsection{SLP under Brezovich criterion conditions}\label{subsect SLP for Brezovich conditions}
First, we will show how small departures from perfect spherical shape affect the SLP. Fig.\ref{Fig_SAR_Bre_r_f's_Victor_Inset} displays the SLP dependence on the elongation ratio $r$ obtained by adding the corresponding shape uniaxial effective anisotropy to the cubic magnetocrystalline anisotropy for particles with the same size as before. All the hysteresis loops were simulated at field amplitudes $H_{max}$ that maximize the Brezovitz criterion at the corresponding frequency.

As can be seen in Fig. \ref{Fig_SAR_Bre_r_f's_Victor_Inset}, even small changes in the particle shape (aspect ratio, \textit{r}) have an important impact on the heating performance. Notice that SLP increases first with increasing $r$, reaching a maximum at a value that depends on the frequency, while for higher aspect ratios it progressively decreases towards zero.
This behavior can be understood noticing that an increase in $r$ means an increase in the uniaxial shape anisotropy (see Eq. \ref{eq:Ku}) added to the cubic crystalline one. Therefore, the effective anisotropy of a spherical particle with crystalline cubic anisotropy will be eventually dominated by the uniaxial shape contribution for sufficiently large aspect ratios, i.e. sufficiently large $K_{sh}$ values.
\begin{figure}[thpb]
   \centering
    \includegraphics[width = \columnwidth]{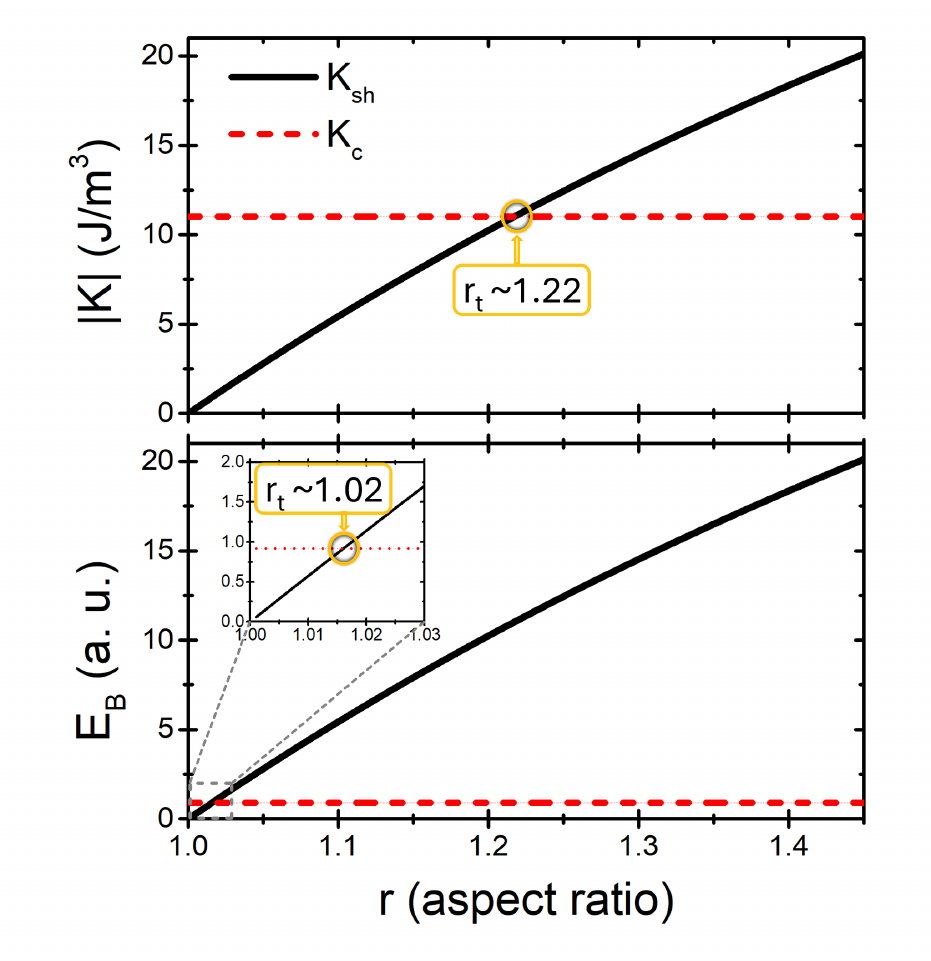}
    \caption{(a) Dependence of the effective uniaxial shape anisotropy $K_{sh}$ on the aspect ratio $r$ of an ellipsoidal particle with major axis $c$ along the $z$ axis. The absolute value of the cubic crystalline anisotropy of magnetite is indicated by the dashed horizontal line. Both become equal at $r_t$. (b) Dependence on the aspect ratio $r$ of the energy barrier per unit volume $E_B$ between the easy-axis direction and the energy maximum (or saddle point) in zero magnetic field.} 
    \label{Fig_Scheme_K_Eb_r}
 \end{figure}

At this point it may be worth emphasising that while heating is defined by the anisotropy, it is not only the anisotropy constant what is important, but also its symmetry. Thus, what matters regarding the hysteresis loops area is the relaxation time, that depends exponentially on the energy barriers separating energy minima and saddle points in the energy landscape induced by the anisotropy. In the case of uniaxial anisotropy, the energy barriers at zero field are simply proportional to the volume $E_{B,u}=K_u V$, but, for cubic anisotropy, the energy barrier separating the minima along the $[111]$ directions from the saddle points along the $[110]$ planes is lower by a factor of $12$: $E_{B,c}=|K_c| V/12$. Therefore, in this case, while the anisotropy constant of the shape contribution becomes bigger than the magnetocrystalline one when the aspect ratio becomes bigger than $r_t=1.22$ [see Fig. \ref{Fig_Scheme_K_Eb_r}(a)], which corresponds to $K_{sh}= 1.1\times 10^4$ J/m$^3$, the shape contribution to the energy barrier dominates over the cubic crystalline at aspect ratios as small as $r_t= 1.02$ [see Fig. \ref{Fig_Scheme_K_Eb_r}(b)]. 

The schemes displayed in Fig. \ref{Fig_Scheme_K_Eb_r} explain why, even if a spherical particle with the considered size would not dissipate significant heat, a slight distortion from its spherical shape will allow it to dissipate in the considered frequency range. The critical importance of this fact is the reason why we decided to explicitly show these otherwise simple schemes.
The maximum SLP (SLP$_{\text{max}}$) is reached for the lowest considered frequency $f= 100$ kHz and the aspect ratio $r_{max}$ that maximizes the SLP varies with the frequency of the AC field.
The value of of SLP$_{\text {max}}$ rapidly decreases with increasing frequencies [see Fig. \ref{Fig_SAR_Bre_r_f's_Victor_Inset} (b)], which can be understood by the fact that the maximum field at high $f$ is so small that the hysteresis loops are almost closed. The same tendency is obtained for the aspect ratio at which the SLP is maximized [see Fig. \ref{Fig_SAR_Bre_r_f's_Victor_Inset} (c)]. These results have important implications for hyperthermia design, as it is suggested that for this particle size, the maximum heating will be obtained by the smaller frequency, at a moderate aspect ratio of $r\simeq 1.12$.

From these observations, we conclude that the best choice for optimization of SLP under physiological limits corresponds to lower frequencies and slightly elongated NP shapes. Increasing the frequency would help if the condition of major loops (fields higher than the anisotropy field) was accomplished, which is clearly not the case if the Brezovich criterion has to be respected.

\begin{figure}[thbp]
   \centering
    \includegraphics[width = \columnwidth]{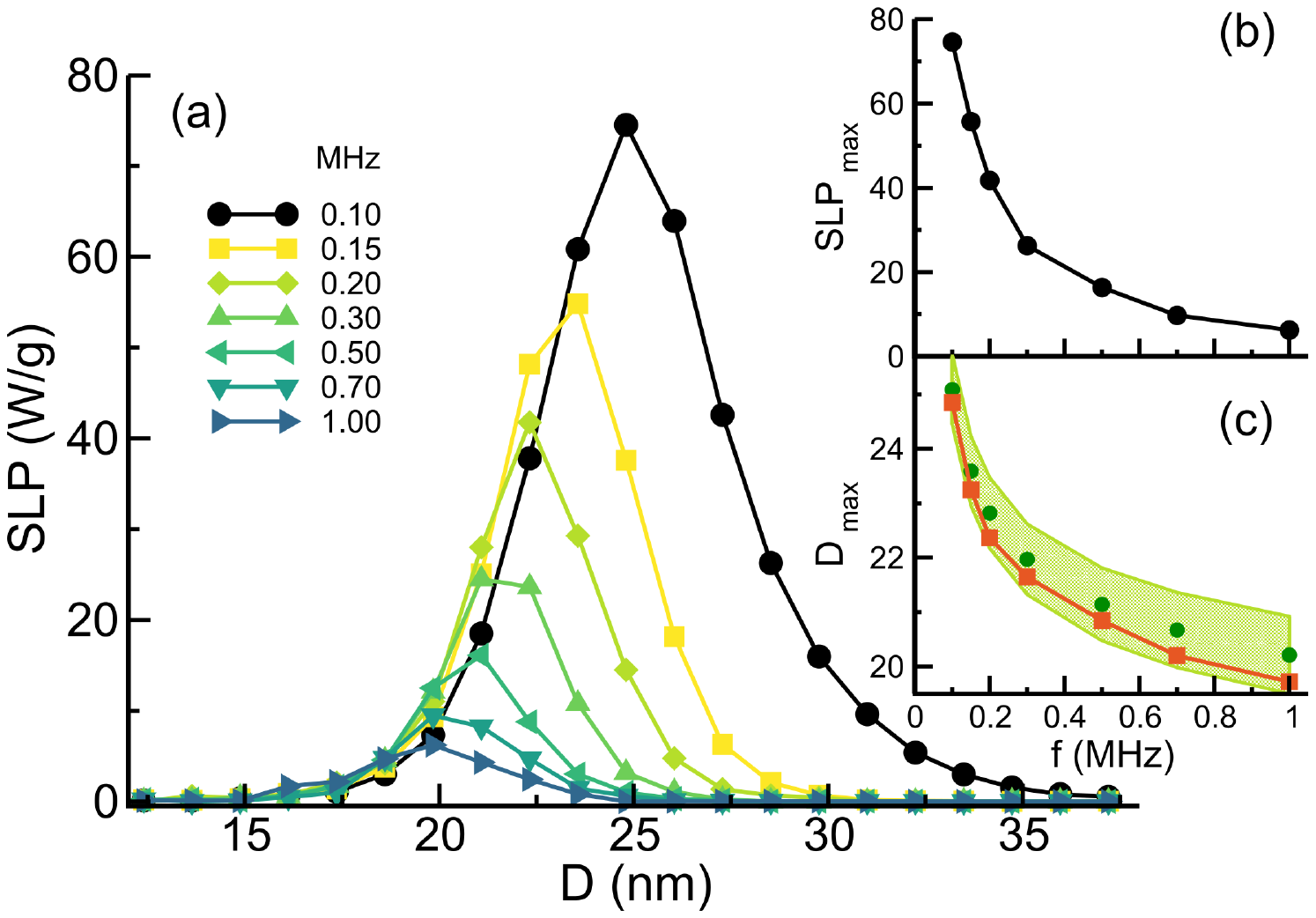}
    \caption{(a) Dependence of the SLP on nanoparticle diameter $D$ and aspect ratio $r=1.12$ ($K_{sh}\simeq 6.5$ kJ/m$^3$) for the frequencies indicated in the legend, and AC fields respecting the Brezovich criterion. A cubic anisotropy constant equal to that of magnetite has been considered. (b) and (c) display the frequency dependence of the maximum value of SLP and the diameter at which is obtained, $D_{max}$. In (c), the green circles are the $D_{max}$ values obtained from Eq. \ref{Eq:Vc} using a value of $f_0=10^9$ Hz and the green shaded region the range obtained when varying $f_0$ between $0.5-1.5\times 10^9$ Hz.   
		} 
    \label{Fig_Brezovich_SAR_vs_D_r1_12_fs}
 \end{figure}

One may wonder if the SLP could also be tuned by varying the particle size for a given value of the aspect ratio $r$. 
For this, we have simulated hysteresis loops at different frequencies again respecting the Brezovich criterion, varying the simulation cell size between $L= 10-30$ nm and keeping $r=1.12$  ($K_{sh}\simeq 6.5$ kJ/m$^3$), which corresponds to particle diameters in the range $D=12.4-37.2$ nm. The corresponding SLP values deduced from the areas of the loops are shown in Fig. \ref{Fig_Brezovich_SAR_vs_D_r1_12_fs}. 

As expected, particles with sizes below $\sim 15$ nm do not heat at any frequency, since they are superparamagnetic at the considered temperature and frequency range. Particles above a certain size do not heat because the associated energy barriers are too high to be overcome by the considered combinations of $H$ and $f$.
At each frequency, there is a particle size $D_{max}$ that optimizes the SLP and it increases as the frequency decreases, whereas the maximum SLP shows the contrary tendency [see insets (b) and (c) of Fig. \ref{Fig_Brezovich_SAR_vs_D_r1_12_fs}], in a way similar to the previously studied dependence of SLP on $r$ in Fig. \ref{Fig_Brezovich_SAR_vs_D_r1_12_fs}. 

We have found that the sizes that maximize the SLP are in very good agreement with the critical size deduced from the Arrhenius-Ne\'el relaxation law for uniaxial anisotropy and randomly oriented axes \cite{Victora_PRL1989}
\begin{equation}
\label{Eq:Vc}
V_c=\frac{k_B T}{K_u}\ln\left(\frac{f_0}{f}\right)\left[1-\frac{H}{H_k}\right]^{-3/2} \ ,
\end{equation}
where $V_c$ is the critical volume and $H_k$ the anisotropy field. Inserting $K_{u}= 6.5\times 10^3$ J/m$^3$, $T= 300$ K, and considering the different frequencies and maximum applied fields, we obtain the values indicated by the green dots in Fig. \ref{Fig_Brezovich_SAR_vs_D_r1_12_fs}(c) when setting $f_0=10^9$ Hz. 
Taking into account an uncertainty of $5\times 10^9$ Hz in the value of $f_0$, the simulation values fall within the acceptance range (marked by the green shaded region), confirming that the optimum particle sizes are related to over-barrier thermal fluctuations.

It is very interesting to observe that varying the size has a very similar frequency-dependence than varying the aspect ratio. This is to say, same that for a given size the highest SLP is obtained for the smaller $f$, also for a given aspect ratio (i.e. anisotropy), the SLP is maximised for smaller $f$. This is a very important aspect to keep in mind, as a common consideration in the literature has been whether a large field or large frequency would be preferred \cite{liu2022field}.

\begin{figure}[thbp]
   \centering
    \includegraphics[width = \columnwidth]{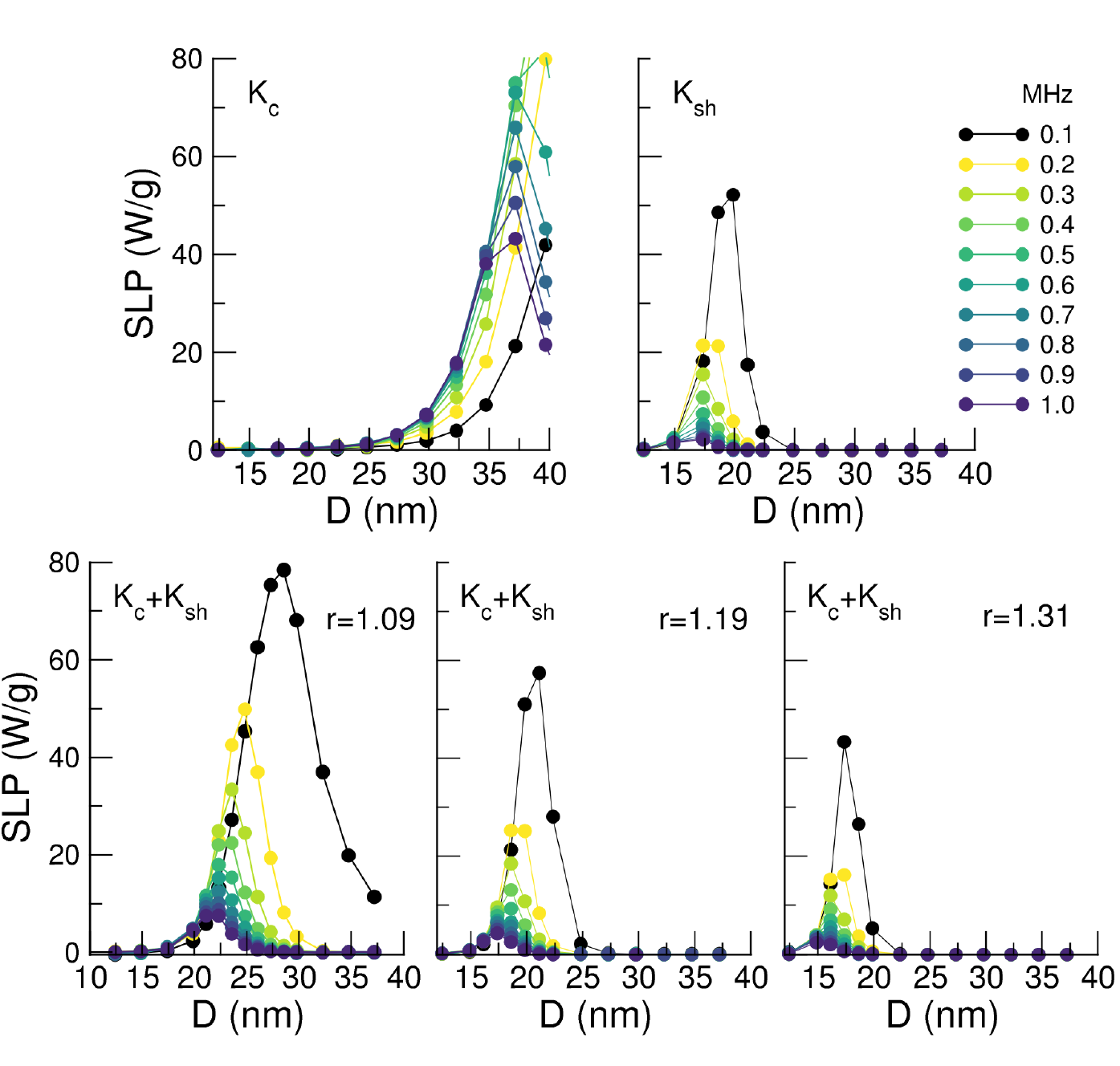}
    \caption{Size-dependence of the SLP for different frequencies and maximum AC fields accomplishing the Brezovich criterion $H_{max}\cdot f = 4.85\times 10^8$ A/m$\cdot$s. Upper panels show the case for spherical particles with only crystalline cubic (left, $K_c$) and only shape (right, with $K_{sh}=K_c$) anisotropies, respectively. Lower panels correspond to ellipsoidal particles of aspect ratios $r= 1.09, 1.19, 1.31$ with uniaxial shape anisotropy added to the cubic one.} 
    \label{Fig_Brezovich_SAR_Bre_D_f's_new}
 \end{figure}

In order to complete the previous description about the influence of the size and shape on the SLP, we will now compare the size dependence of SLP of particles having only cubic anisotropy (corresponding to spherical shape), only uniaxial shape anisotropy (usual approach in the literature for spherical particles) and ellipsoidal particles with different aspect ratios. The results of the simulations obtained again for a wide range of frequencies and fields matching the Brezovich criterion for the physiological conditions are presented in Fig. \ref{Fig_Brezovich_SAR_Bre_D_f's_new}, for the only cubic and only unixial cases (top panels, left and right, respectively), and for cubic plus uniaxial cases, for aspect ratios $r= 1.09, 1.19, 1.31$ corresponding to shape anisotropies $K_{sh}= 5, 10, 15$ kJ/m$^3$, respectively (bottom panels Fig. \ref{Fig_Brezovich_SAR_Bre_D_f's_new}, from left to right).

\begin{figure}[htbp]
   \centering
    \includegraphics[width = 0.8\columnwidth]{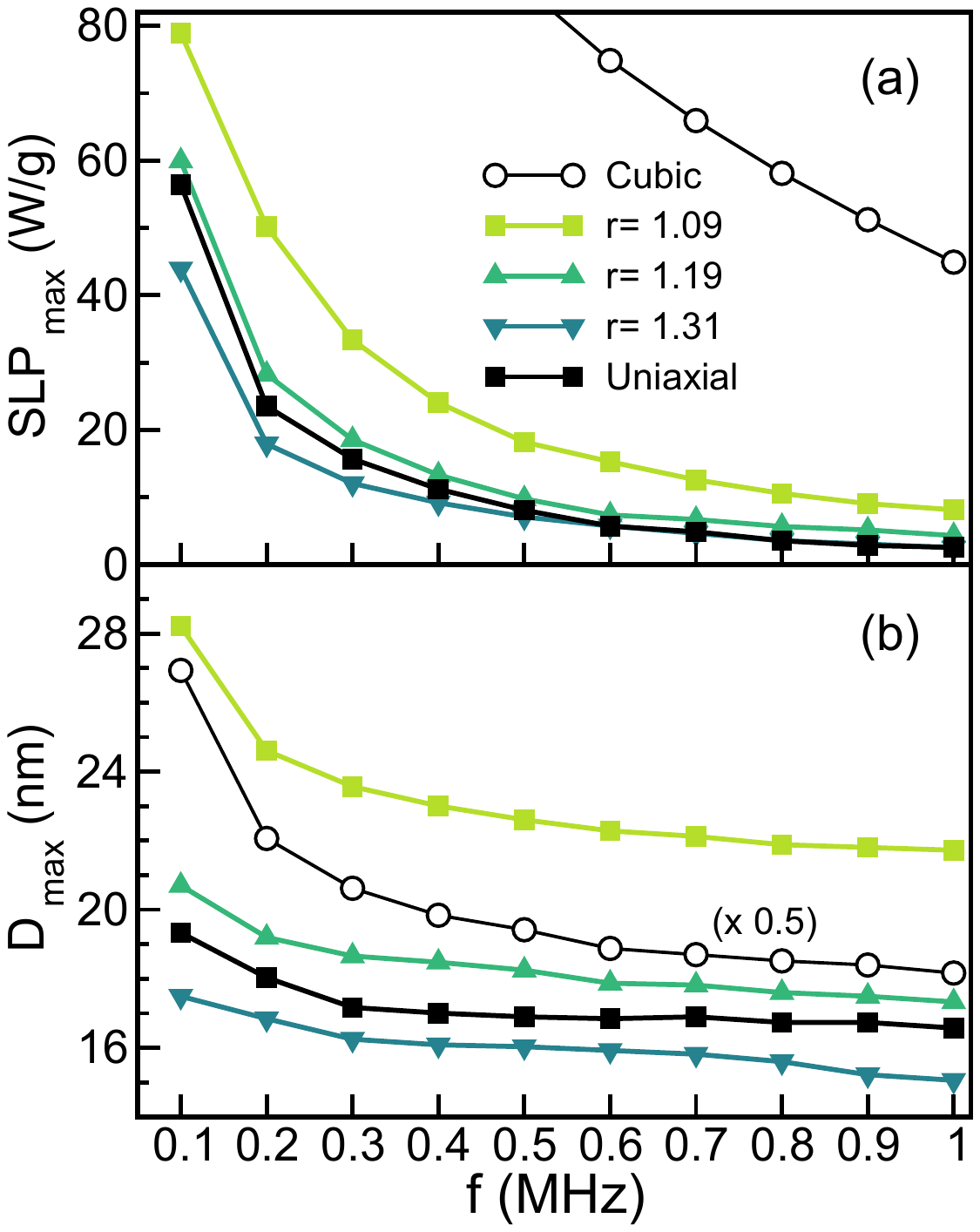}
    \caption{Frequency dependence of the (a) maximum SLP and (b) corresponding particle size extracted from the curves in Fig. \ref{Fig_Brezovich_SAR_Bre_D_f's_new} for different nanoparticle elongation ratios $r= 1.09, 1.19, 1.31$ (colored symbols) and also for the cases of only uniaxial or only (black squares) cubic contribution (empty circles) with $K_{u,c}= 1.1\times 10^4$ J/m$^3$. Notice that $D_{max}$ for the last case has been rescaled by $0.5$.}
    \label{Fig_Brezovich_Dmax_SARmax}
 \end{figure}

The variation of the heating properties as displayed in Fig. \ref{Fig_Brezovich_SAR_Bre_D_f's_new} is complex but presents some systematic characteristics.
First, notice that the SLP is always maximized for a certain particle size for all the frequencies and aspect ratios.
SLP attains maximum values always at the lowest studied frequency $f= 100$ kHz, which corresponds to $H_{max}=6.1$ mT, independently of the model used for the anisotropy and decreases rapidly with increasing frequency for all $r$. This is accompanied by a reduction of the optimum NP diameter $D_{max}$, as can be see in Fig. \ref{Fig_Brezovich_Dmax_SARmax}(b).

Secondly, the values of $SLP_{max}$ are considerably lowered even for moderate increases of the aspect ratio, while the optimal sizes $D_{max}$ are more moderately reduced by $r$. 
This can be understood as follows: an increase of the aspect ratio increases the uniaxial shape anisotropy that, for NPs with relatively high values of $M_s$ dominates over the crystalline cubic contribution already for aspect ratios as low as $r=1.02$ (see Fig. \ref{Fig_Scheme_K_Eb_r}); therefore, in order to have appreciable  magnetization reversal probability, the particle volume has to be reduced so as to have sizeable energy losses.  
We also observe that the window of NP sizes around $D_{max}$ that give sizable SLP values is reduced when increasing $r$, which is related again to the increase of the effective shape anisotropy that translates into more pronounced changes of the energy barriers with volume.  
 
Next, we compare these results for elongated NPs to the case of spherical ones with crystalline cubic anisotropy only, to show the critical role played by the magnetocrystalline contribution. The results are shown in the upper left panel of Fig.\ref{Fig_Brezovich_SAR_Bre_D_f's_new}. Notice that, although for the lowest frequencies the peaks are not visible in the represented scale, they range from $D_{max}= 54$ to $D_{max}= 36$ nm as shown in Fig. \ref{Fig_Brezovich_Dmax_SARmax}(a) (empty circles).
We observe that the maximum SLP of NPs with cubic anisotropy increases by a factor that varies between $\sim 2$ at $f= 0.1$ MHz ($133$ W/g) and $\sim 8$ at $f=1.0$ MHz ($45$ W/g) when compared to the SLP$_{\text{max}}$ of NPs with only $K_u$. 
Moreover, the sizes for optimal heating performance are also more than doubled as compared to those of the only uniaxial anisotropy case [see the black squares in Fig. \ref{Fig_Brezovich_Dmax_SARmax}(b); notice the reduction factor used for the cubic case].  
The reason for this has to be traced back to the above mentioned reduction in the anisotropy energy barriers by a factor of $12$ as compared to uniaxial anisotropy. According to Eq. \eqref{Eq:Vc}, this decrease is translated into an increase of the critical volume and in a higher SLP$_{\text{max}}$ when the real cubic anisotropy of the NP is included. Nevertheless, it is important to remember that this ideal case is only shown for theoretical purposes, as experimentally it is not possible to synthesize particles with such a small sphericity; for the sake of comparison, experimental data reporting "spherical particles" is usually considered for aspect ratios between 1.05 and 1.10.


\subsection{Interaction effects}\label{subsect interactions}
So far, we have not taken into account dipolar interparticle interactions; this is a reasonable first approach for theoretical considerations, but is only a valid assumption for highly dispersed NP ensembles. Different degrees of dilution and interparticle distances can be somehow regulated exvivo when NP are still dissolved in water by employing different coatings that adhere to the NP surface and thus avoiding close proximity between them.  
However, when administered to biological media, NPs tend to aggregate forming clusters or agglomerates whose size and spacial distribution cannot be deliberately controlled \cite{Beola_ACSAppl2020}. Moreover, in order to achieve significant local heating, high enough NP doses have to be administered to the tumor and, when they are internalized by cells, their aggregation may be forced by the reduced volume of the vesicles in which they may be contained  \cite{Sanz_SciRep2016}. It is thus crucial to understand how interactions may affect heating production in physiological conditions. 

Neither experimental nor theoretical published works on the subject have reached a consensus regarding the effect of dipolar interactions on SLP. 
The disparity of the results stems from the fact that depending on the spatial NP arrangement and orientation of their easy-axes, dipolar interactions can increase or decrease the energy barriers responsible for magnetization reversal, thus affecting their hysteretical properties.
In spite of the variety of results, a model based on simulations of NPs with different intrinsic magnetic features, magnetic field conditions and concentrations, showed that the variety of conflicting heat dissipation results can be actually described by a single picture \cite{conde2015single}.
Moreover, several works have demonstrated that a considerable increase of heating power can be achieved in spite of interactions by NP chain formation either favoured by cubic shaped NPs \cite{Martinez-Boubeta_SciRep2013}, promoted by the applied AC field \cite{serantes2014multiplying} along the applied field direction, or naturally present in magnetosomes \cite{Gandia_Small2019,Fdez-Gubieda_JAP2020,Marcano_JPCC2020}. 
In contrast, the variety of results in assemblies with no particular spatial order arises from the fact that, while in some cases samples may contain diluted suspensions of individual aggregates or small NP clusters uniformly distributed in space, others are formed by larger multicore NP aggregates \cite{Gavilan_ChemSocRev_2021}. 

\begin{figure}[tbph]
   \centering
    \includegraphics[width = \columnwidth]{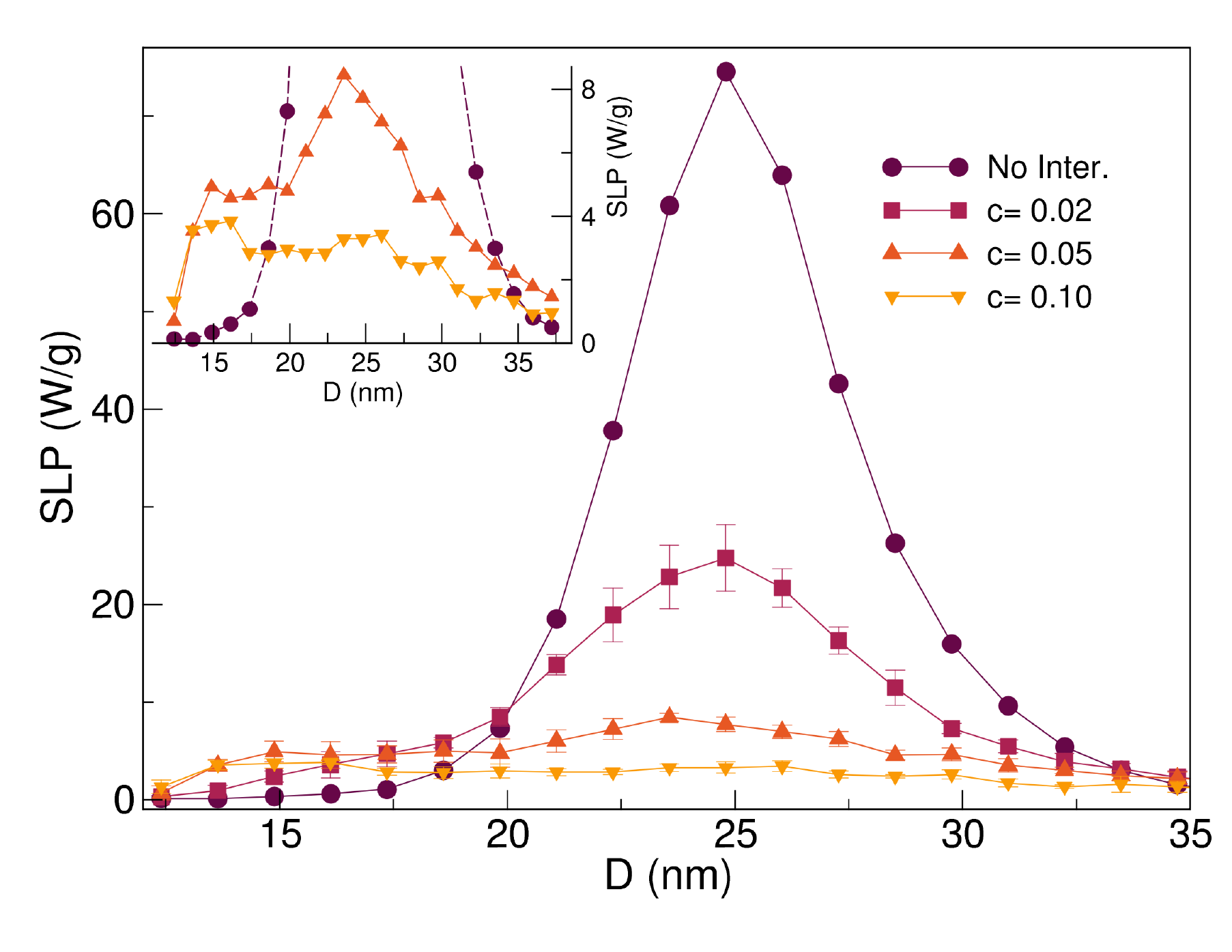}
    \caption{Size-dependence of SLP under Brezovich criterion for a frequency $100$ kHz and different volume concentrations $c= 2 \%, 5 \%, 10 \%$ for an ellipsoidal particle with aspect ratio $r=1.12$. Error bars indicate the dispersion after averaging over $4$ independent runs. The inset shows a zoomed in region where the SLP size dependence for the two higher concentrations can be more clearly discerned. 
    }
    \label{Fig_SAR_Bre_D_r1_12_f100_Inter}
 \end{figure}

In what follows, we present simulations of hysteresis loops based on ensembles of NPs randomly distributed in space with different volume concentrations, prepared following the strategy presented in Sec. \ref{Sec_Simulation}. We will focus on ellipsoidal NPs with $r= 1.12$ under an $f= 100$ kHz AC field, since they gave the highest SLP in the non-interacting case.
The results of the SLP extracted from the simulated hysteresis loops after averaging over $4$ independent runs are presented in Fig. \ref{Fig_SAR_Bre_D_r1_12_f100_Inter} for volume concentrations $c= 2 \%, 5 \%, 10 \%$. Notice that due to the intrinsic limitations of the OOMMF code, more diluted diluted samples were not considered since simulation times become too long due to the large atlas sizes that have to be considered (OOMMF deals with all simulation cells in the atlas, even those not assigned to a NP).
Compared to the hysteresis loops of the non-interacting case, those of interacting NPs, become more tilted with a considerable decrease of the remanent magnetization and a decrease of the magnetization at the maximum field. Although the coercive field does not present such an appreciable reduction, the overall result on the HL is a pronounced decrease of the $SLP_{max}$, that is reduced from $75$ W/g to $25$ W/g for a concentration $c= 2 \%$ and to values below $10$ W/g for higher $c$. 
A similar reduction of the SLP has also been reported in a work by Gubanova \textit{et al.} \cite{gubanova2021heating} under different AC field conditions.
Although, the NP size optimizing heating seems not to be much affected by the interaction, being around $D_{max}= 25$ nm, the window of dissipating sizes becomes broader, extending to smaller sizes as $c$ is increased. 
Looking at the zoomed in inset of Fig.\ref{Fig_SAR_Bre_D_r1_12_f100_Inter}, it is worth highlighting that the interparticle interactions seem to activate the heat release of small sized NPs that did not heat in the absence of interactions, while leaving unaffected the big ones that did not dissipate in the non-interacting case. This is an indication that dipolar interactions act to increase the effective energy barriers of otherwise superparamagnetic NPs at room temperature. 
This has important consequences when aiming at controlling precisely the SLP in real samples which will always have some degree of polydispersion. 

We speculate that the drastic reduction of SLP values even for moderate concentrations, although may seem surprising at first sight, could be due to local non-homogeneities of the NP positions. 
Although, as explained in Sec.\ref{model}, the NPs were distributed at random inside a cube, we have detected the presence of NPs in close contact for all the studied concentrations, being more numerous for increasing $c$. 
In particular, the percentage of touching NPs is $6.6 \%$ for $c= 2 \%$ and, for $c= 10 \%$, the number increases up to $26.2 \%$. 
We have ascertained by visual inspection of the NP spatial distributions that most of the cases correspond to dimers, although we cannot exclude the existence of a smaller fraction of trimers in the most concentrated case. 
Since the interactions are most important for these clusters, even the presence of a small fraction of them could influence significantly the global heating behavior of the assembly, as reported elsewhere \cite{Niculaes2017}. 
The results of a study about the heating properties of small NP clusters by Ortega-Julia et al. \cite{Ortega-Julia_Nanoscale2023} also indicate this fact. 

Hysteresis loops at $f= 1000$ kHz (not shown) were also simulated and become very narrow and elongated, giving much lower values of SLP that do not show any discernible trend with $D$ due to noisiness. The interactions completely suppress any heating at $c= 10\%$.

%
%
%
%

\section{Discussion and Conclusions}\label{conclusions}

A key general conclusion of our work is related to the choice of field amplitude/frequency combination to be used in hyperthermia treatments: the results suggest that, independently on the anisotropy model used to describe the particles, the best is to use the highest $H_{max}$ value as possible (equivalently, the smallest $f$). This is a rather general result that we think deserves further experimental test. 

In relation to the one of the main objectives of the work, that was to study the applicability of the usual uniaxial-only anisotropy to describe hyperthermia performance under the Brezovich criterion conditions, we have shown that the magnetocrystalline contribution definitively plays a key role. This is to say, the usual uniaxial-only magnetic anisotropy approximation seems quite limited to guide/understand \textit{in vivo} biological conditions for hyperthermia applications. This is also very relevant from the application viewpoint, as it is directly related to the role of the NP shape: even very small deviations in shape can have an important impact on the heating performance. This is also an important factor to take into account towards the standardization of magnetic hyperthermia as a clinical protocol \cite{beola2019roadmap}.

Regarding the role of dipolar interactions on the SLP, we have shown that in ensembles of NPs randomly distributed in space, even concentrations as small as $2\%$ hinder the approach to saturation causing a prominent decrease in the hysteresis loops area and heating performance, even at the optimum NP size and field conditions obeying the Brezovich criterion.
This degradation of the heating properties, can be partially mitigated in practice by covering the NPs with surfactants to avoid the close contact between them and the formation of clusters. 

Finally, it is also important to stress the fact that all results presented here correspond to the ideal "frozen ferrofluid" assumption, i.e. that the particles do not move under the action of the external AC field. While such has been a common theoretical assumption for AC fields with frequencies $\ge{100}$ kHz as considered in this work \cite{usov2010low}, it has been also theoretically pointed out that a strong reorientation effect could take place at very small frequencies \cite{Mamiya_SciRep2011}. 
More work is necessary to clarify this matter, both from the theory \cite{Usov_Jap2012} and experimental \cite{wang2023probing} points of view, as it is known that magnetic NPs in living systems may adopt several spatial configurations in biological environments, in general intermediate between the ideal frozen ferrofluid and the fully movable situations. For example, particles may be internalized within endosomes or lysosomes, resulting in agglomerated arrangements \cite{Beola_ACSAppl2020,Rojas2017}; may have an extracellular location, for example located within the extracellular matrix with different aggregation degrees \cite{Beola2021}; or may even exhibit either quite diluted distributions or chain-like ones depending on the particle compositions (even for very similar samples \cite{Balakrishnan2020}).
Furthermore, it would also be important to consider the possible reorientation of the interacting NPs with respect to the AC field \cite{Gubanova_Beilstein2021}, with the role of the dynamical colloidal evolution \textit{per se}, as it has been shown that particles may form elongated structures under the application of the AC field \cite{saville2014formation,morales2021time}. And, it is well known that chaining may strongly change the heating performance \cite{Martinez-Boubeta_SciRep2013,serantes2014multiplying,Orue_Nanoscale2018,Gandia_Small2019,Marcano_ACSNano2022}. 
The possibility that chain-like NP formation could push the SLP$_{\text{max}}$ to values similar to those for the non-interacting case under Brezovich conditions, is an interesting alternative that we are planning to study in forthcoming studies.


\section*{Conflicts of interest}
There are no conflicts to declare.

\section*{Acknowledgements}
We acknowledge financial support by Spanish Ministerio de Ciencia, Innovaci\'on y Universidades through projects PID2019-109514RJ-I00 and PID2021-127397NB-I00, "ERDF A way of making Europe", by the "European Union", and Catalan DURSI (2021SGR0032). Xunta de Galicia is acknowledged for project ED431F 2022/005. AEI is also acknowledged for  the \textit{Ram\'on y Cajal} grant RYC2020-029822-I that supports the work of D.S. We acknowledge the Centro de Supercomputacion de Galicia (CESGA) for computational resources.

\balance

\providecommand*{\mcitethebibliography}{\thebibliography}
\csname @ifundefined\endcsname{endmcitethebibliography}
{\let\endmcitethebibliography\endthebibliography}{}

\end{document}